\documentclass[pre,preprint,showpacs,eqsecnum]{revtex4}
\usepackage{graphicx}
\begin{document}
\title{Ratchet-transport for a chain of interacting charged
particles}
\author{S. I. Denisov}
\email{denisov@sumdu.edu.ua}
\author{E. S. Denisova}
\affiliation{Department of Mechanics and Mathematics, Sumy State
University, 2, Rimskiy-Korsakov Street, 40007 Sumy, Ukraine}
\author{Peter H\"{a}nggi}
\email{hanggi@physik.uni-augsburg.de} \affiliation{Institut
f\"{u}r Physik, Universit\"{a}t Augsburg,
Universit\"{a}tsstra{\ss}e 1, D-86135 Augsburg, Germany}

\begin{abstract}
We study analytically and numerically the overdamped,
deterministic dynamics of a chain of {\it charged}, interacting
particles driven by a longitudinal alternating electric field and
additionally interacting with a smooth ratchet potential. We
derive the equations of motion, analyze the general properties of
their solutions and find the drift criterion for chain motion. For
ratchet potentials of the form of a double-sine and a
phase-modulated sine it is demonstrated that both, a so-called
integer and fractional transport of the chain can occur. Explicit
results for the directed chain transport for these two classes of
ratchet potentials are presented.
\end{abstract}

\maketitle

\section{Introduction}

During the last decade, much attention was devoted to the study of
the so-called ratchet effect, i.e., the rectifying of
nondirectional stochastic or deterministic driving forces into
directional motion of (quasi-)particles, for  comprehensive
reviews, see, for example, Refs.~\cite{R02,Linke02,JAP97,
A97,AH02}. This phenomenon is of prominent practical importance
and constitutes a theoretical basis for Brownian motors
\cite{A97,AH02}, particle segregation \cite{RSAP94,FBKL95},
reduction of a vortex density \cite{LJDB99}, smoothing of a
crystalline surface \cite{Kehr1997,DLB98,Miretartes}, and many
others \cite{R02,Linke02,AH02}.

There exists a large variety of systems that exhibit the ratchet
effect. Those of them where transported particles are exposed to a
driving force and move in an asymmetric periodic (ratchet)
potential form a class of so-called rocking ratchets
\cite{Bartussek1994}. In turn, stochastic and deterministic
rocking ratchets are distinguished, depending on whether they are
driven only deterministically, or whether they are subjected also
to stochastic forces. The fluctuating noise component  in turn
allows for activated escape events \cite{HTB} even at
sub-threshold driving. The stochastic situation has been studied
for both, noninteracting \cite{Bartussek1994,M93,Bartussek1996}
and interacting \cite{CFV97,ZHH98, Wambaugh1999, Savelev2004}
particles, and the strong influence of the inter-particle
interactions was revealed. In contrast, the deterministic case has
been investigated primarily for noninteracting particles
\cite{JKH96,SL99,M00,S01,BCM02}. At variance with earlier work
\cite{Marchesoni1996} focusing on the transport of driven linear
defects (i.e. elastic chains) which diffuse on asymmetric
substrates at finite temperatures, the objective of this study is
to investigate the transport properties of a chain of
\textit{charged}, interacting particles in deterministic rocking
ratchets. This model can be used for the investigation of
one-dimensional crystals in carbon nanotubes \cite{Hirahara00} and
ion transport through synthetic nanopores \cite{SF02}.

The paper is organized as follows. In Sec.~II, we derive the
reduced system of two equations that describes the deterministic,
overdamped motion of a chain. We carry out the general analysis of
this system in Sec.~III. Specifically, we introduce a class of its
steady-state solutions, analyze the chain dynamics in the inverted
potential, and derive the drift criterion of a chain. In Sec.~IV,
we study analytically and numerically the overdamped transport of
a chain in the double-sine ratchet potential. Concluding remarks
are contained in Sec.~V, and the phase-modulated sine potential is
introduced in the Appendix.

\section{Equations of motion}

We study a chain of charged particles of identical masses $M$ that
interact among each other via the Coulomb interaction, a repulsive
interaction, and additionally with a (substrate) ratchet potential
$V(x)$ having the period $d$. Moreover, the particles are driven
by a longitudinal alternating electric field $E(t)$ of a temporal
period $2T$. Therefore, the total potential energy of a chain
$V_{\mathrm{tot}}$ includes the interaction energy $V_{\mathrm
{int}}$ and the potential energies $V_{\mathrm{r}}$ and
$V_{\mathrm{el}}$ formed by a ratchet potential and an electric
field, respectively. Assuming that any neighboring particles have
opposite charges, i.e., $q_{i+1} = -q_{i}$ and $|q_{i}|=q$, and a
repulsive interaction depends on the inter-particle distance as
$|x_{i}-x_{j}|^{-r}$ [$x_{i}=x_{i}(t)$ is the coordinate of the
$i$th particle, $r>1$ since it prevents the unbounded contraction
of a chain], these three parts of the total energy can be written
as follows:
\begin{eqnarray}
    V_{\mathrm{int}} =
    \frac{1}{2}{\sum_{i,j}}'\frac{q_{i}q_{j}}{|x_{i}- x_{j}|}
    +\frac{b}{2}{\sum_{i,j}}'\frac{1}{|x_{i}-x_{j}|^{r}}\,,
    \nonumber\\
    V_{\mathrm{r}}=\sum_{i}V(x_{i}), \quad
    V_{\mathrm{el}} = -\sum_{i}q_{i}x_{i}E(t).\phantom{a}
    \label{eq:energies}
\end{eqnarray}
Here $b$ is a dimensioned parameter characterizing the strength of
the repulsive interaction, and the primes on the summation signs
imply that $i\neq j$.

Along with the potential force $-\partial V_{\mathrm{tot}}/
\partial x_{i}$ on the $i$th particle a friction force $-\lambda
\dot{x}_{i}$ ($\lambda$ is a damping coefficient) is also acting.
Then the equation of motion of each, individual particle reads
\begin{equation}
    M\ddot{x}_{i}+\lambda\dot{x}_{i}+\frac{\partial
    V_{\mathrm{int}}}
    {\partial x_{i}}=q_{i}E(t)+f(x_{i}),
    \label{eq:eq_motion1}
\end{equation}
where $f(x)=-V'(x)$ (here and below, the prime denotes the
derivative with respect to the argument of the function) is a
force field generated by a ratchet potential. The set of equations
(\ref{eq:eq_motion1}) is very complicated for a detailed analysis
of the chain motion because they are coupled and contain the
nonlinear terms $\partial V_{\mathrm {int}}/ \partial x_{i}$ and
$f(x_{i})$. Nevertheless, the problem of coupled ratchets does
possess a very rich behavior, such as anomalous hysteresis,
self-oscillations, absolute negative mobility, etc.
\cite{Reimann1999}, and thus is demanding to analyze. Since our
aim is to study a ratchet mechanism of the chain transport, the
nonlinear nature of the ratchet force $f(x_{i})$ must be taken
into account, while the nonlinearity of the interaction force
$\partial V_{\mathrm {int}}/\partial x_{i}$ seemingly is not so
essential. Therefore, in order to partially simplify the problem,
we restrict ourselves to a harmonic approximation for
$V_{\mathrm{int}}$. This approximation  is valid if during the
period $2T$ of the action of the electric field $E(t)$ the
particle displacements are much less than the equilibrium distance
$a$ [at $V(x)=0$ and $E(t)=0$] between the nearest particles.

According to the findings in \cite{DD03}, a periodic chain with
equidistant particles exists only if $r>r_{2} \approx 2.799$; in
the opposite case the minimum of $V_{\mathrm{int}}$ occurs for a
chain with infinite period. Assuming that the condition $r>r_{2}$
holds, it is convenient to introduce the coordinates $2na+x_{n}^
{+}$ and $(2n+1)a+x_{n}^{-}$ [$n$ ($n=0,\pm1,...$) numbers the
chain cells which contain two particles (see Fig.~\ref{fig1})] of
positive and negative charges, respectively. The equations of
motion for the displacements $x_{n}^{+}=x_{n}^{+}(t)$ and
$x_{n}^{-} = x_{n} ^{-}(t)$ of these particles from the
equilibrium positions follow from Eq.~(\ref{eq:eq_motion1}). Using
the harmonic approximation for $V_{\mathrm{int}}$ \cite{DD03} and
the notations $E(t) = Eh(t)$ and $f(x)=f_{0}g(x)$, where $E$ is
the amplitude of $E(t)$, $f_{0}= |\min{f(x)}|$, and $h(t)$ and
$g(x)$ are the dimensionless electric and ratchet force fields,
respectively, the equations of motion assume the form
\begin{eqnarray}
    \lefteqn{\ddot{x}_{n}^{+} + 2\Omega_{\lambda} \dot{x}_{n}^{+}
    + \Omega^{2}\sum_{m}\tilde{B}_{2(n-m)-1} (x_{n}^{+} -
    x_{m}^{-}) {}} \nonumber\\
    && {} + \Omega^{2}\sum_{m\neq n}\tilde{B}_{2(n-m)}(x_{n}^{+}
    - x_{m}^{+})=Ah(t)+Rg(x_{n}^{+}),\nonumber\\
    \lefteqn{\ddot{x}_{n}^{-} + 2\Omega_{\lambda}
    \dot{x}_{n}^{-} + \Omega^{2}\sum_{m}\tilde{B}_{2(n-m)+1}
    (x_{n}^{-} - x_{m}^{+}) {}} \\
    && {} + \Omega^{2}\sum_{m\neq n}\tilde{B}_{2(n-m)}(x_{n}^{-}
    - x_{m}^{-})=-Ah(t)+Rg(x_{n}^{-}).\nonumber
    \label{eq:eq_motion2}
\end{eqnarray}
Here $\Omega_{\lambda}=\lambda/2M$, $\Omega^{2}=q^{2}/Ml^{3}$,
$l=(b/q^{2})^{1/(r-1)}$ is the length scale, $A=qE/M$,
$R=f_{0}/M$,
\begin{equation}
    \tilde{B}_{n}=\frac{1}{\gamma_{1}^{3}(r)}\left[2
    \frac{(-1)^{n}}{|n|^{3}}+\frac{(r+1)\ln2}{\zeta(r)
    |n|^{r+2}}\right],
    \label{eq:tildeB}
\end{equation}
$\gamma_{1}(r)= [r\zeta(r)/\ln2]^{1/(r-1)}$, $a=l\gamma_{1}(r)$,
and $\zeta(r)= \sum_{n=1}^{\infty} n^{-r}$ is the Riemann zeta
function. Note that Eqs.~$(2.3)$ are valid if the condition
$|x_{n}^{\pm} - x_{m}^{\pm}|\ll a$ holds for all $n$ and $m$. It
provides the applicability of the harmonic approximation and, as
can be shown readily, it does not prohibit the existence of
unbounded (at $t\to\infty$) solutions of these equations that
describe the chain transport.

The system of equations $(2.3)$ constitutes a theoretical basis
for the study of ratchet-transport of ionic chains within the
harmonic approximation for the inter-particle interactions.
Unfortunately, at present there are no methods to find and analyze
its unbounded solutions. Since this system contains an infinite
number of coupled nonlinear equations, the development of such
methods is a very complicated problem which, in general, can be
solved only approximately. However, we will show that, in a
particular case, the chain transport can be studied in detail
without approximations.

More specifically, we consider the case when the equilibrium chain
period $a$ in absence of the ratchet potential $V(x)$ is a whole
integer of the potential period $d$, i.e., $a=Ld$, where $L$ is a
natural number. Although this commensurability assumption is
rather restrictive (small discrepancy between $a$ and $Ld$
violates the chain periodicity in presence of the ratchet
potential), it permits us to reduce the infinite system $(2.3)$ to
the system of only two equations. Indeed, if the initial
conditions $x_{n}^ {\pm}(0) = x_{\pm}$ and $\dot{x}_{n}^{\pm}(0)=
v_{\pm}$ ($x_{\pm}$ and $v_{\pm}$ are the initial displacements
and velocities of the positive and negative charges) hold for all
$n$, then all positively charged particles and all negatively
charged particles move identically. Designating in this case
$x_{n}^{\pm} = x^{\pm}$, from Eqs.~$(2.3)$ we get
\begin{equation}
    \begin{array}{c}
    \displaystyle \ddot{x}^{+} + 2\Omega_{\lambda} \dot{x}^{+} +
    \mbox{$\frac{1}{2}$}\omega^{2}(1)({x}^{+} - {x}^{-})
    = Ah(t)+Rg(x^{+}), \\[8pt]
    \displaystyle \ddot{x}^{-} + 2\Omega_{\lambda} \dot{x}^{-} +
    \mbox{$\frac{1}{2}$}\omega^{2}(1)({x}^{-} - {x}^{+})
    = -Ah(t)+Rg(x^{-}),
    \end{array}
    \label{eq:eq_motion2_red}
\end{equation}
where $\omega^{2}(1) = 2\Omega^{2} \sum_{m} \tilde{B}_{2m-1}$ is
the squared frequency of optical vibrations of chain particles
that corresponds to the dimensionless wave number $\kappa=1$
\cite{DD03}. Note that Eqs.~(\ref{eq:eq_motion2_red}) are derived
from Eqs.~$(2.3)$ without approximations and they precisely
describe the chain dynamics in the examined case. However, since a
nonlinear oscillator driven by a periodic force can exhibit
chaotic behavior that is characterized by strong sensitivity to
initial conditions \cite{Moon}, the chain dynamics governed by
Eqs.~$(2.3)$ and (\ref{eq:eq_motion2_red}) can be quite different
if the corresponding initial conditions slightly differ. With
increasing damping constant the chaotic domain in the parameter
space is reduced, therefore we expect (and this is confirmed by
simulations) that it vanishes in the overdamped limit
($\Omega_{\lambda} \to \infty$). In other words, the chain
dynamics in this limit is expected to be regular and predictable.

Introducing the dimensionless time $\tau=t/2T$ and the
dimensionless particle displacements $w \equiv w(\tau)=
x^{+}(2T\tau)/d$ and $u \equiv u(\tau)=x^{-}(2T\tau)/d$, from
Eqs.~(\ref{eq:eq_motion2_red}) we obtain in the overdamped limit
\begin{equation}
    \begin{array}{c}
    \displaystyle \chi\frac{dw}{d\tau}=\phi H(\tau)-
    \mbox{$\frac{1}{2}$}(w-u) + \mu G(w),\\[8pt]
    \displaystyle \chi\frac{du}{d\tau}=-\phi H(\tau)-
    \mbox{$\frac{1}{2}$}(u-w) + \mu G(u).
    \end{array}
    \label{eq:eq_motion3}
\end{equation}
Here $\chi = \Omega_{\lambda}/T\omega^{2}(1)$ and $\phi =
A/d\omega^{2}(1)$ are the dimensionless parameters characterizing
the electric field frequency and amplitude, respectively, $H(\tau)
= h(2T\tau)$, $\mu = R/d\omega^{2}(1)$ is the dimensionless
parameter characterizing the ratchet force amplitude, $G(\bar{x})
= g(d\bar{x})$, $\bar{x}=x/d$ is the dimensionless coordinate, and
according to Eq.~(\ref{eq:tildeB}){\setlength\arraycolsep{2pt}
\begin{eqnarray}
    \omega^{2}(1) &=& \Omega^{2}\bigg(\frac{\ln2}{r\zeta(r)}
    \bigg)^\frac{3}{r-1}\bigg[4\ln2\,(1-2^{-r-2})(r+1)
    \nonumber\\[4pt]
    &&  \times\frac{\zeta(r+2)}{\zeta(r)} - 7\zeta(3)\bigg].
    \label{eq:omega(1)}
\end{eqnarray}}Assuming that all particles at $t=0$ are in
equilibrium (their equilibrium positions are those in absence of
the ratchet potential), the initial conditions for
Eqs.~(\ref{eq:eq_motion3}) are written as $w(0)=u(0)=0$. The
system of equations (\ref{eq:eq_motion3}) provides a very useful
tool for studying the transport properties of ionic chains.
Indeed, on the one hand, it is a rather simple set of only two
coupled, ordinary differential equations of first order and, on
the other hand, it accounts for the inter-particle interactions
and the action of the ratchet force.

By the definitions, the functions $H(\tau)$ and $G(\bar{x})$ have
unit periods, $H(\tau+1)= H(\tau)$ and $G(\bar{x}+1) =
G(\bar{x})$, and zero mean values, $\int_{0}^{1} H(\tau)d\tau=0$
and $\int_{0}^{1} G(\bar{x}) d\bar{x}=0$. The latter condition
shows that the total work, delivered by the ratchet force field
$f(x)$ on any interval of length $d$, equals zero. In general, the
functions $H(\tau)$ and $G(\bar{x})$ can be both, continuous and
discontinuous. But here, to simplify the numerical solution of
Eqs.~(\ref{eq:eq_motion3}), we consider them as smooth,
differentiable functions.

In what follows, to find the drift criterion of a chain, we shall
use the dimensionless potential energy $U=U(w,u,\tau)$ that
reduces Eqs.~(\ref{eq:eq_motion3}) to the form
\begin{equation}
    \chi\frac{dw}{d\tau}= - \frac{\partial}{\partial w}U,
    \qquad
    \chi\frac{du}{d\tau}= - \frac{\partial}{\partial u}U.
    \label{eq:eq_motion4}
\end{equation}
Introducing the representation $V(x)=V_{0}W(\bar{x})$, where
$V_{0}>0$ and $W(\bar{x})$ is the dimensionless ratchet potential,
and by use of the definitions of $f(x)$ and $f_{0}$, we get
$f_{0}=(V_{0}/d)\max{W'(\bar{x})}$ and
\begin{equation}
    G(\bar{x}) = - W'(\bar{x})/\max{W'(\bar{x})}.
    \label{eq:def_G}
\end{equation}
Finally, from the inspection of the right-hand sides of
Eqs.~(\ref{eq:eq_motion3}) and (\ref{eq:eq_motion4}) we obtain
{\setlength\arraycolsep{2pt}
\begin{eqnarray}
    U &=& - \phi\sin(2\pi\tau)(w-u) + \mbox{$\frac{1}{4}$}(w-u)^2
    \nonumber\\[4pt]
    && + \mu[W(w) + W(u)]/\max{W'(\bar{x})}.
    \label{eq:def_U}
\end{eqnarray}

\section{General Results}

\subsection{Periodicity analysis}

In order to exclude from consideration transient processes, we
need to examine the asymptotic, steady-state solutions of
Eqs.~(\ref{eq:eq_motion3}). These solutions depend on many
factors, such as the form of a ratchet potential, characteristics
of a chain, initial conditions, etc., and consequently can be
studied in detail only numerically. However, using the periodicity
of $H(\tau)$ and $G(\bar{x})$, it is possible to introduce
different classes of the steady-state solutions. One particular
such class is generated by those periodic solutions of
Eqs.~(\ref{eq:eq_motion3}) that asymptotically $\tau\to\infty$
obey:
\begin{equation}
    w(\tau+k)=w(\tau)+K, \quad u(\tau+k)=u(\tau)+K,
    \label{eq:sym1}
\end{equation}
where $k$ and $K$ are natural and integer numbers, respectively,
that have no common factors. In this case, the periodicity and
increment of the functions $w(\tau)$ and $u(\tau)$ are described
by the pair $\{k,K\}$ which at $K \neq 0$ corresponds to the drift
state of a chain. Since the reduced chain displacement $\Delta w =
\lim_{\tau\to\infty}[w(\tau + k) - w(\tau)]/k$ that occurs during
one period of $H(\tau)$ is given by $\Delta w = K/k$ (note that
the harmonic approximation is valid if $|K|/k \ll L$), we shall
term the chain transport characterized by the pair $\{k,K\}$ as
``integer'' if $k=1$, and ``fractional'' if $k \geq 2$.

According to the conditions (\ref{eq:sym1}), the average velocity
of a chain or drift velocity $v=(d/2T) \lim _{\tau \to \infty}
w(\tau) /\tau \, [=(d/2T)\lim_{\tau\to \infty} u(\tau)/ \tau]$ is
reduced to
\begin{equation}
    v=\frac{d}{2T}\lim_{\tau\to\infty}\frac{w(\tau+k)-w(\tau)}{k}
    \label{eq:vel1}
\end{equation}
which, in turn, yields $v=v_{0}\bar{v}$, where $v_{0}=d
\omega^{2}(1) /2\Omega_{\lambda}$ and $\bar{v} = \chi K/k$ is the
dimensionless drift velocity. Taking into account that the
periodicity and drift parameters $k$ and $K$ depend, in general,
on all parameters of Eqs.~(\ref{eq:eq_motion3}), we conclude that
$\bar{v}$ is a discontinuous linear function of $\chi$.

We emphasize that this class does not exhaust all the steady-state
solutions of Eqs.~(\ref{eq:eq_motion3}). Moreover, the symmetry
approach does not permit us to find their unique steady-state
solution in each concrete case. Therefore, to study the transport
properties of a chain, it is necessary to numerically find the
solution of Eqs.~(\ref{eq:eq_motion3}) (with zero-valued initial
conditions) and examine its long-time behavior depending on the
form of a ratchet potential, the electric field characteristics,
and the chain parameters.

\subsection{Chain dynamics in the inverted potential}

We now consider the chain dynamics in the inverted potential
$V_{\mathrm{in}}(x) = V(-x)$ that generates the reduced force
field $G_{\mathrm{in}}(\bar{x})$. According to
Eqs.~(\ref{eq:eq_motion3}), in such a potential the displacements
$w_{\mathrm{in}}(\tau)$ and $u_{\mathrm{in}}(\tau)$ of positively
and negatively charged particles from their equilibrium positions
are governed by the equations of motion
\begin{equation}
    \begin{array}{c}
    \displaystyle\chi\frac{dw_{\mathrm{in}}}{d\tau}=\phi
    H(\tau)-\mbox{$\frac{1}{2}$}(w_{\mathrm{in}}-u_{\mathrm{in}})
    +\mu G_{\mathrm{in}}(w_{\mathrm{in}}),\\[8pt]
    \displaystyle\chi\frac{du_{\mathrm{in}}}{d\tau}= - \phi
    H(\tau)-\mbox{$\frac{1}{2}$}(u_{\mathrm{in}}-w_{\mathrm{in}})
    +\mu G_{\mathrm{in}}(u_{\mathrm{in}})
    \end{array}
    \label{eq:eq_motion5}
\end{equation}
[$w_{\mathrm{in}}(0)=u_{\mathrm{in}}(0)=0$]. Taking into account
that $G_{\mathrm{in}}(-\bar{x})= -G(\bar{x})$, one can show from
Eqs.~(\ref{eq:eq_motion3}) and (\ref{eq:eq_motion5}) that
$w_{\mathrm{in}}(\tau) = -u(\tau)$ and $u_{\mathrm{in}}(\tau) =
-w(\tau)$. This implies that if the chain dynamics is known for
the ratchet potential $V(x)$, then it is also known for the
corresponding inverted potential $V_{\mathrm{in}}(x)$ as well. In
particular, if a chain in the potential $V(x)$ drifts along the
$x$-axis, then in the inverted potential $V_{\mathrm{in}}(x)$ it
drifts in the opposite direction with the same average velocity,
i.e., $v_{\mathrm{in}} = -v$. We emphasize that in
reflection-symmetric potentials the drift state of a chain, i.e.,
a chain state with $v\neq0$, does not exist. Indeed, as it is
shown above, the general condition $v_{\mathrm{in}}=-v$ must hold.
On the other hand, if $V(-x) = V(x)$ then $V_{\mathrm{in}}(x) =
V(x)$ and so the condition $v_{\mathrm{in}}=v$ also must hold. It
is obvious that both conditions are met simultaneously only if
$v=0$.

The above mentioned features of the chain dynamics permit us to
study the overdamped transport of ionic chains only in those
ratchet potentials that induce the chain drift, say, with positive
velocity $v$. In the following we consider the simple-structured
ratchet potentials. We assume that the reduced potentials
$W(\bar{x})$ and the corresponding force fields $G(\bar{x})$ have
only one maximum and one minimum on unit period and, in addition,
$\max{G(\bar{x})}>1$, see Fig.~\ref{fig2}.

\subsection{Drift criterion}

To find the conditions that lead to the drift state of a chain, we
rewrite Eqs.~(\ref{eq:eq_motion3}) as
\begin{equation}
    \begin{array}{c}
    \displaystyle \chi\frac{dw}{d\tau} = \phi H(\tau) -
    \mbox{$\frac{1}{2}$}(w-u) + \mu G(w),\\[8pt]
    \displaystyle \chi\frac{d}{d\tau}(w+u) = \mu[G(w)+G(u)],
    \end{array}
    \label{eq:eq_motion6}
\end{equation}
where the second equation is obtained by summing
(\ref{eq:eq_motion3}). If $\mu/\chi \to 0$, then it reduces to the
equation $d(w+u)/d\tau=0$, which with $w(0)=u(0)=0$ yields
$w(\tau)= -u(\tau)$. Using this relation and the condition that
$\mu/\chi \to 0$, the first equation in (\ref{eq:eq_motion6})
takes the form $\chi dw/d\tau + w = \phi H(\tau)$. Its solution
\begin{equation}
    w(\tau) = \frac{\phi}{\chi}\int_{0}^{\tau}H(\tau-\tau')
    e^{-\tau'/\chi}d\tau'
    \label{eq:solut1}
\end{equation}
shows that
\begin{equation}
    w(\tau+1) = w(\tau) + \frac{\phi}{\chi}e^{-\tau/\chi}
    \int_{0}^{1}H(-\tau')e^{-\tau'/\chi}d\tau',
    \label{eq:relat1}
\end{equation}
and so $w(\tau+1) = w(\tau)$ for $\tau\to\infty$. This means that
the drift state of a chain does not exist if $\mu/\chi \to 0$.
Since the parameter $\chi$ is proportional to the electric field
frequency, its decrease leads to an increase of the maximal
particle displacements, yielding $v|_{\chi>0} = 0$  if
$v|_{\chi=0} = 0$. The last condition is violated if the amplitude
parameter $\phi$ is large enough. Therefore, to find the drift
criterion of a chain, we need to consider its dynamics as
$\chi\to0$.

According to Eqs.~(\ref{eq:eq_motion6}), in the stationary regime
($\chi\to0$) the chain dynamics is described by the system of
nonlinear equations
\begin{equation}
    \begin{array}{c}
    \phi H(\tau) - \mbox{$\frac{1}{2}$}(w-u) + \mu G(w)=0,
    \\[6pt]
    G(w)+G(u)=0.
    \end{array}
    \label{eq:eq_motion7}
\end{equation}
If for each time the chain energy $U$ has a minimum value, i.e.,
if in virtue of Eqs.~(\ref{eq:eq_motion7}) the conditions
\begin{equation}
    \frac{\partial^{2}U}{\partial w^{2}}>0, \quad
    \frac{\partial^{2}U}{\partial w^{2}}
    \frac{\partial^{2}U}{\partial u^{2}}-
    \left(\frac{\partial^{2}U}{\partial w\partial u}
    \right)^{2}>0
    \label{eq:cond2}
\end{equation}
hold, then $v|_{\chi=0}=0$. The latter condition in
(\ref{eq:cond2}), $G'(w) + G'(u) - 2\mu G'(w)G'(u) < 0$, is weaker
than the former, $2\mu G'(w) -1 < 0$. Hence, it is violated first
under increasing of the parameter $\phi$. Let $\phi_{\mathrm{cr}}$
be the critical value of the parameter $\phi$ such that, for $\phi
= \phi_ {\mathrm{cr}}$, the latter inequality in (\ref{eq:cond2})
is reduced to equality at some instant of time. It is obvious that
this occurs for the first time at $\tau = \tau_{1}$, where
$\tau_{1} (<1)$ is the minimal solution of the equation
$H(\tau_{1})=1$. Then, using the first equation in
(\ref{eq:eq_motion7}), we obtain
\begin{equation}
    \phi_{\mathrm{cr}} = \mbox{$\frac{1}{2}$}(w_{\mathrm{cr}}^{-}
    -u_{\mathrm{cr}}^{-}) - \mu G(w_{\mathrm{cr}}^{-}),
    \label{eq:drift_crit}
\end{equation}
where $w_{\mathrm{cr}}^{-} = w(\tau_{1}-0)$ and $u_{\mathrm{cr}}
^{-} = u(\tau_{1}-0)$ are defined by the system of equations
\begin{equation}
    \begin{array}{c}
    G'(w_{\mathrm{cr}}^{-})+G'(u_{\mathrm{cr}}^{-})-
    2\mu G'(w_{\mathrm{cr}}^{-})G'(u_{\mathrm{cr}}^{-}) = 0,
    \\[6pt]
    G(w_{\mathrm{cr}}^{-})+G(u_{\mathrm{cr}}^{-})=0.
    \end{array}
    \label{eq:cond3}
\end{equation}

At $\phi = \phi_ {\mathrm{cr}}$ and $\tau=\tau_{1}+0$ the chain
particles instantly move to the new equilibrium positions
$w_{\mathrm{cr}}^{+} = w(\tau_{1}+0)$ and $u_{\mathrm{cr}}^{+} =
u(\tau_{1}+0)$, which are defined by another system of equations
\begin{equation}
    \begin{array}{c}
    \phi_{\mathrm{cr}} - \mbox{$\frac{1}{2}$}(w_{\mathrm{cr}}^{+}
    -u_{\mathrm{cr}}^{+}) + \mu G(w_{\mathrm{cr}}^{+})=0,
    \\[6pt]
    G(w_{\mathrm{cr}}^{+})+G(u_{\mathrm{cr}}^{+})=0.
    \end{array}
    \label{eq:cond4}
\end{equation}
As Fig.~\ref{fig2} illustrates, in this case the positively
charged particles pass into the next potential wells, while the
negatively charged particles do not leave their own wells. A
detailed analysis shows that during the second and each following
periods of $H(\tau)$ both types of particles instantly move into
the next potential wells twice. In other words, during each period
of an alternating electric field a chain in the steady-state
regime is displaced by two periods of a ratchet potential.

Thus, the drift criterion of a chain, that leads to the condition
$v|_{\chi=0} \neq 0$, has the form $\phi > \phi_{\mathrm{cr}}$. As
$\phi \to \phi_{\mathrm{cr}}$ and $\chi \to 0$ a chain exhibits
the integer transport with the drift parameter $K=2$ and with an
average velocity $\bar{v} = 2\chi$. According to
Eqs.~(\ref{eq:drift_crit}) and (\ref{eq:cond3}), to calculate
$\phi_{\mathrm{cr}}$ it is necessary to know the explicit form of
a ratchet potential. However, taking into account that for
$w=\bar{x}_{1}$ and $u=\bar{x}_{5}-1$ the conditions $G(w)+G(u)=0$
and $v|_{\chi=0} = 0$ hold, we find the general condition
\begin{equation}
    \phi_{\mathrm{cr}} > \mbox{$\frac{1}{2}$}(\bar{x}_{1} -
    \bar{x}_{5} + 1) + \mu,
    \label{eq:eval}
\end{equation}
which can be used for approximate estimation of
$\phi_{\mathrm{cr}}$.

Note also that, because for noninteracting particles the chain
energy (\ref{eq:def_U}) does not contain the term $(w-u)^{2}/4$,
the drift criterion of free particles assumes the form $\phi >
\mu$. The main distinction between the drift states of interacting
and noninteracting particles thus is the result that in the latter
case $K\to\infty$ as $\chi\to0$.

\section{Double-sine potential}

\subsection{Analytical results}

We examine the chain dynamics in the asymmetric  ratchet potential
composed of two spatial harmonics \cite{Bartussek1994}. This
so-called double-sine potential is defined as $V_{\mathrm{d}}(x) =
V_{0\mathrm{d}} W_{\mathrm{d}} (\bar{x})$, where $V_{0\mathrm{d}}$
is a positive constant and
\begin{equation}
    W_{\mathrm{d}}(\bar{x})=-\sin[2\pi(\bar{x}+\bar{x}_{\mathrm{d}})]
    -\eta_{\mathrm{d}}\sin[4\pi(\bar{x}+\bar{x}_{\mathrm{d}})].
    \label{eq:potential1}
\end{equation}
Here, the parameter $\eta_{\mathrm{d}}(>0)$ characterizes the form
of the reduced potential $W_{\mathrm{d}}(\bar{x})$, and the
parameter $\bar{x}_{\mathrm{d}}=x_{\mathrm{d}}/d$ defines the
positions of its extrema. The function $W_{\mathrm{d}}(\bar{x})$
and the corresponding reduced force field $G_{\mathrm{d}}
(\bar{x}) = -W_{\mathrm{d}}'(\bar{x})/ \max{W_{\mathrm{d}}'
(\bar{x})}$ [$f_{\mathrm{d}}(x) = f_{\mathrm{0d}}G_{\mathrm{d}}
(\bar{x})$, $f_{\mathrm{0d}} = (V_{\mathrm{0d}}/d) \max{W
_{\mathrm{d}}' (\bar{x})}$] both have qualitatively the same forms
as those depicted in Fig.~\ref{fig2}, i.e., they have only two
extrema per unit period and $\min{W_{\mathrm{d}} (\bar{x})} =
W_{\mathrm{d}} (0)$, if $\eta_{\mathrm{d}} < \eta_{0\mathrm{d}} =
1/8$ and
\begin{equation}
    \bar{x}_{\mathrm{d}} = \frac{1}{2\pi}\arccos
    \frac{\sqrt{1+32\eta_{\mathrm{d}}^{2}}-1}
    {8\eta_{\mathrm{d}}}.
    \label{eq:def_xd}
\end{equation}
(Yet another ratchet potential that possesses the same properties
as $V_{\mathrm{d}}(x)$ is introduced in the Appendix.) For these
conditions, we find that $\max{W_{\mathrm{d}}' (\bar{x})} =
2\pi(1-2\eta_{\mathrm{d}})$, $\min{W_{\mathrm{d}}' (\bar{x})} =
-2\pi(1+2\eta_{\mathrm{d}})$, $f_{\mathrm{0d}} =
2\pi(1-2\eta_{\mathrm{d}})V_{\mathrm{0d}}/d$,
{\setlength\arraycolsep{2pt}
\begin{eqnarray}
    G_{\mathrm{d}}(\bar{x}) &=& \{
    \cos[2\pi(\bar{x}+\bar{x}_{\mathrm{d}})]
    +2\eta_{\mathrm{d}}\cos[4\pi(\bar{x}+\bar{x}_
    {\mathrm{d}})]\}    \nonumber\\[4pt]
    && \times(1-2\eta_{\mathrm{d}})^{-1},
    \label{eq:def_Gd}
\end{eqnarray}
$G_{\mathrm{d}} (\bar{x}_{1}) = \min{G_{\mathrm{d}} (\bar{x})} =
-1$ if $\bar{x}_{1} = 1/2-\bar{x}_{\mathrm{d}}$, $G_{\mathrm{d}}
(\bar{x}_{2}) = G_{\mathrm{d}} (0) = 0$ if $\bar{x}_{2} =
1-2\bar{x}_{\mathrm{d}}$, $G_{\mathrm{d}} (\bar{x}_{4})=
\max{G_{\mathrm{d}} (\bar{x})} = (1+2\eta_{\mathrm{d}}) /(1-2\eta_
{\mathrm{d}})$ if $\bar{x}_{4} = 1-\bar{x}_{\mathrm{d}}$, and
$G_{\mathrm{d}} (\bar{x}_{3}) = G_{\mathrm{d}} (\bar{x}_{5}) = 1$
if
\begin{equation}
    \bar{x}_{3,5} = 1-\bar{x}_{\mathrm{d}} \mp
    \frac{1}{2\pi}\arccos\frac{\sqrt{1+16\eta_{\mathrm{d}}}-1}
    {8\eta_{\mathrm{d}}},
    \label{eq:def_x3,5}
\end{equation}
where the upper and lower signs correspond to the indexes 3 and 5,
respectively.

To calculate $\phi_{\mathrm{cr}}$, we proceed as follows. First,
instead of the second equation in (\ref{eq:cond3}), we introduce
the two equations $G(w_{\mathrm{cr}}^{-}) = -\rho$ and $G(u_{
\mathrm{cr}}^{-}) = \rho$ ($-1 \leq \rho \leq 1$). Then, taking
into account the conditions $\bar{x}_{1} < w_{\mathrm{cr}} ^{-} <
\bar{x}_{3}$ and $\bar{x}_{5} - 1 < u_{\mathrm{cr}}^{-} <
\bar{x}_{1}$, we find their solutions
\begin{equation}
    {w_{\mathrm{cr}}^{-} \choose u_{\mathrm{cr}}^{-}} =
    \frac{1\pm1}{2} - \bar{x}_{\mathrm{d}} \mp \frac{1}{2\pi}
    \arccos\frac{Z(\pm\rho)-1}{8\eta_{\mathrm{d}}},
    \label{eq:sol_w_u}
\end{equation}
where $Z(\rho)=\sqrt{1 - 16\eta_{\mathrm{d}}\rho +
32\eta_{\mathrm{d}}^{2}(1 + \rho)}$, and, using
Eqs.~(\ref{eq:def_Gd}) and (\ref{eq:sol_w_u}), we reduce the first
equation in (\ref{eq:cond3}) to the form
\begin{equation}
    \sum_{\sigma}\frac{\sigma\eta_{\mathrm{d}}(1-
    2\eta_{\mathrm{d}})}{Z(\sigma\rho)\sqrt{64
    \eta_{\mathrm{d}}^{2}-[Z(\sigma\rho)-1]^{2}}}
    = \frac{\pi}{2}\mu
    \label{eq:eq_rho}
\end{equation}
($\sigma=\pm1$). Since the left-hand side of
Eq.~(\ref{eq:eq_rho}), $L(\rho)$, is a monotonic odd function and
$L(\rho)\to\infty$ as $\rho\to1$, this equation always has a
unique solution with respect to $\rho$. If that solution is known,
then from Eqs.~(\ref{eq:drift_crit}) and (\ref{eq:sol_w_u}) we get
the desired formula
\begin{equation}
    \phi_{\mathrm{cr}} = \frac{1}{2} + \mu\rho -\frac{1}{4\pi}
    \sum_{\sigma}\arccos\frac{Z(\sigma\rho)-1}
    {8\eta_{\mathrm{d}}}.
    \label{eq:drift_crit2}
\end{equation}
According to Eqs.~(\ref{eq:eq_rho}) and (\ref{eq:drift_crit2}),
$\phi_{\mathrm{cr}}$ is a universal function of $\mu$ and $\eta_
{\mathrm{d}}$. A corresponding 3-D plot, obtained via the
numerical solution of Eq.~(\ref{eq:eq_rho}), shows, see
Fig.~\ref{fig3},  that $\phi_ {\mathrm{cr}}$ is an almost linear
function of these variables.

An analytical solution of Eq.~(\ref{eq:eq_rho}) is possible only
in some special cases. Specifically, if $\eta_{\mathrm{d}}\to0$
then, calculating the leading term of $L(\rho)$,
Eq.~(\ref{eq:eq_rho}) gives
\begin{equation}
    \frac{\rho(3-2\rho^{2})}{(1-\rho^{2})^{3/2}}
    = \pi\frac{\mu}{\eta_{\mathrm{d}}}.
    \label{eq:eq_rho2}
\end{equation}
At $\mu/\eta_{\mathrm{d}}\to\infty$ its approximate solution reads
$\rho = 1- (\eta_{\mathrm{d}}/\pi\mu)^{2/3}/2$, and
Eq.~(\ref{eq:drift_crit2}) yields $\phi_{\mathrm{cr}} = 1/4 +
\mu$. Another example corresponds to the limit $\mu \to 0$.
Because $L(0)=0$, the solution of Eq.~(\ref{eq:eq_rho}) tends to
zero as $\mu \to 0$, and so $\phi_{\mathrm{cr}} = 1/2 - \bar{x}_
{\mathrm{d}}$. We emphasize that the cases $\eta_{\mathrm{d}}=0$
and $\mu=0$ are degenerate. This means that for
$\eta_{\mathrm{d}}=0$ and $\mu=0$ the drift state of a chain never
exists, while for $\eta_{\mathrm{d}}\neq0$ and $\mu\neq0$ it is
realizable. The reason lies in the breakdown of the spatial
symmetry.

\subsection{Numerical results}

We solved Eqs.~(\ref{eq:eq_motion3}) with zero initial conditions
by the fourth-order Runge-Kutta method for $H(\tau)=
\sin(2\pi\tau)$ and $G(\bar{x}) = G_{\mathrm{d}}(\bar{x})$. The
analysis shows that each steady-state solution of
Eqs.~(\ref{eq:eq_motion3}) satisfies the conditions
(\ref{eq:sym1}). At $\phi > \phi_ {\mathrm{cr}}$, the typical
dependence of the dimensionless chain displacement $\Delta w$ on
$\chi$ is depicted in Fig.~\ref{fig4}~a. The changes of $\Delta w$
occur in a  very narrow intervals $(\chi_{1}^{s}, \chi_{2}^{s})$
(the index $s$ labels these intervals) of the $\chi$-axis. We
found that if the parameter $\chi$ does not belong to these
intervals, then $k=1$, and the chain dynamics is characterized by
the pairs $\{1,K\}$. According to our terminology, a chain
exhibits an integer transport if $K \neq 0$. Its main features are
as follows. First, the chain velocity $\bar{v} = K\chi$ is a
piecewise linear function of $\chi$ that has a number of local
maxima, see Fig.~\ref{fig4}~b. These maxima occur due to the
discrete character of the drift parameter $K$. Second,
$K|_{\chi=0}$ is an increasing, step-like function of $\phi$ that
equals zero if $\phi < \phi_{\mathrm{cr}}$ and takes even numbers
if $\phi > \phi_ {\mathrm{cr}}$. Specifically, in accordance with
the analytical results, $K|_{\chi=0} = 2$ at $\phi \approx
\phi_{\mathrm{cr}}$. To illustrate for $\chi << 1$ the chain
dynamics in the drift state, the time dependence of the particle
displacements $w(\tau)$ and $u(\tau)$ and the phase trajectory of
the chain motion are shown in Figs.~\ref{fig5} and \ref{fig6} for
$\phi = 3$ and $\mu = 2$. Since a chain reaches the steady state
only if $\tau >> \chi$ (the relaxation time is of the order of
$\chi$), the functions $w(\tau)$ and $u(\tau)$ represent the
steady-state solution of Eqs.~(\ref{eq:eq_motion3}), which satisfy
the conditions $w(\tau + 1) = w(\tau) + 4$ and $u(\tau + 1) =
u(\tau) + 4$, at $\tau \gtrsim 1$. Finally, the increase of $\chi$
leads to the stepwise decrease of $K$ and a smoothing of $w(\tau)$
and $u(\tau)$.

Solving Eqs.~(\ref{eq:eq_motion3}) for $\chi \in (\chi_{1}^{s},
\chi_{2}^{s})$, we discovered that the chain dynamics is
characterized by the pairs $\{k,K\}$ with $k \geq 2$ and $K \neq
0$. In this case a fractional transport of a chain is realized.
The number of the intervals $(\chi_{1}^{s}, \chi_{2}^{s})$ equals
$K|_{\chi=0}$ ($s=1,2,...,K|_{\chi=0}$) and their width grows with
$s$. Within each such interval the chain displacement $\Delta w =
K/k$ assumes a stepwise function of $\chi$ that satisfies the
condition $K_{s} -1 < K/k < K_{s}$, where $K_{s} = K|_{\chi=0} - s
+ 1$ is the drift parameter to the left of the interval
$(\chi_{1}^{s}, \chi_{2}^{s})$. If $\chi$ approaches its
boundaries on the inside, then $k$ is strongly increased, $K/k \to
K_{s}$ as $\chi \to \chi_{1}^{s}$, and $K/k \to K_{s}-1$ as $\chi
\to \chi_{2}^{s}$. Table~1 illustrates these properties for the
case represented in Fig.~\ref{fig4} at $\chi \in (\chi_{1}^{4},
\chi_{2}^{4})$, where $\chi_{1}^{4} \approx 0.4693877$ and
$\chi_{2}^{4} \approx 0.4732787$. To illustrate the chain dynamics
at $\chi \in (\chi_{1}^{4}, \chi_{2}^{4})$, the time-dependence of
the particle displacements $w(\tau)$ and $u(\tau)$ and the phase
trajectory of the chain motion are depicted in Figs.~\ref{fig7}
and \ref{fig8}, respectively. If $\chi > \chi_{2}^{4}$ then a
drift of the chain does not exist and the phase trajectory
explores a finite region of the phase plane.
\begin{table}
    \caption{The numbers $k$ and $K$ vs $\chi$ for $\chi\in
    (\chi_{1}^{4},\chi_{2}^{4})$.}
    \label{tab:tab}
    \begin{ruledtabular}
    \begin{tabular}{lll|lll}
    $\chi$ & \it{k} & \it{K} & $\chi$ & \it{k} & \it{K} \\
    \hline
    0.46938773 & 135 & 134 & 0.472 & 2 & 1 \\
    0.4693878 & 39 & 38 & 0.473 & 9 & 2 \\
    0.469388 & 21 & 20 & 0.4732 & 8 & 1 \\
    0.46939 & 7 & 6 & 0.47327 & 47 & 2 \\
    0.4694 & 3 & 2 & 0.473278 & 82 & 1 \\
    0.47 & 3 & 2 & 0.4732785 & 150 & 1
    \end{tabular}
    \end{ruledtabular}
\end{table}

The fractional transport occurs also for noninteracting particles.
But the $\chi$-intervals, where such a transport exists, are much
more narrow as compared to those observed in the interacting case.
In particular, according to the above results $\chi_{2}^{4} -
\chi_{1}^{4} \approx 3.9\times10^{-3}$, while for noninteracting
particles the width of the corresponding interval approximately
equals $1.4\times10^{-4}$.

Note that in the phase-modulated sine potential, introduced in the
Appendix, the ionic chain exhibits qualitatively the same
features. We expect, therefore, that the results of the overdamped
ionic chain dynamics are typical and robust for the considered
class of ratchet potentials.

\section{Conclusions}

We have investigated the overdamped transport of a chain of
charged, interacting particles driven by a longitudinal
alternating electric field that additionally interact with a
smooth, non-symmetric, periodic ratchet potential. Assuming that
the equilibrium particle positions coincide with the minima of a
ratchet potential, we have reduced the infinite system of
equations that describes the dynamics of each chain to the system
of two equations which effectively describe the dynamics of only
two, positively and negatively charged, representative particles.
The reduced system  of equations (\ref{eq:eq_motion3}) has the
advantage of being particularly simple because it consists of two
ordinary differential equations of first order, which are driven
by the external force.

Using the time-periodicity of an alternating electric field and
the space-periodicity of a ratchet potential, we have introduced a
wide class of corresponding steady-state solutions of
Eqs.~(\ref{eq:eq_motion3}). The mathematical structure corresponds
to the drift state of the ionic  chain. Particularly, all the
steady-state solutions of Eqs.~(\ref{eq:eq_motion3}) obtained
numerically in the cases of the double-sine and phase-modulated
sine potentials belong to this class. Studying the chain dynamics
in the original ratchet potential and in its inverted realization,
we have shown that, depending on the parameter regime, a chain
either does not drift at all for both realizations, or it has a
finite drift velocity $v$ which is opposite in value, and
correspondingly $-v$, for the inverted ratchet potential.
Considering the chain dynamics in the stationary regime, we have
derived the drift criterion of a chain. Accordingly, the drift
state of a chain takes place if during the first half-period of an
alternating electric field the chain particles perform stick-slip
transitions.

Applying analytical and computational methods for analysis of the
chain dynamics in the double-sine and phase-modulated sine ratchet
potentials, we have shown that the chain displacement, which
occurs during one full period of an alternating electric field, is
a monotonically decreasing, stepwise function of the electric
field frequency. This function, scaled by the ratchet potential
period, takes on only integer and fractional values. Therefore,
only two types of the chain transport, namely integer and
fractional, do exist. Both types occur for tailored frequency
intervals; the frequency intervals, however, that correspond to
the fractional transport are a much more narrow than those
corresponding to the integer transport.

\section*{ACKNOWLEDGMENTS}

We are grateful to T. V. Lyutyy for help with our numerical
calculations. E.S.D. acknowledges the support by INTAS, grant
03-55-1180, and P.H. the support by the Deutsche
Forschungsgemeinschaft, grant HA 1517/13-4.

\appendix*
\section{Phase-modulated sine potential}

We define a new ratchet potential $V_{\mathrm{p}}(x)$, which we
call the phase-modulated sine potential, as $V_{\mathrm{p}}(x) =
V_{0\mathrm{p}} W_{\mathrm{p}} (\bar{x})$, where $V_{0\mathrm{p}}$
is a positive constant and
\begin{equation}
    W_{\mathrm{p}}(\bar{x})=-\sin\{2\pi(\bar{x}+\bar{x}_{\mathrm{p}})
    +\eta_{\mathrm{p}}\sin[2\pi(\bar{x}+\bar{x}_{\mathrm{p}})]\}.
    \label{eq:potential2}
\end{equation}
The dimensionless potential $W_{\mathrm{p}} (\bar{x})$ and the
corresponding dimensionless force $G_{\mathrm{p}} (\bar{x}) =
-W_{\mathrm{p}}'(\bar{x})/ \max{W_{\mathrm{p}}' (\bar{x})}$
[$f_{\mathrm{p}}(x) = f_{\mathrm{0p}}G_{\mathrm{p}} (\bar{x})$,
$f_{\mathrm{0p}} = (V_{\mathrm{0p}}/d) \max{W _{\mathrm{p}}'
(\bar{x})}$] both have only two extrema on the unit period if the
phase amplitude $\eta_{\mathrm{p}}(>0)$ satisfies the condition
$\eta_{\mathrm{p}} < \eta_{0\mathrm{p}} \approx 0.31767 $, and
$\min{W_{\mathrm{p}} (\bar{x})} = W_{\mathrm{p}}(0)$ if the
parameter $\bar{x}_ {\mathrm{p}}$ [$\bar{x}_{\mathrm{p}} \in
(0,1/4)$] is a solution of the equation
\begin{equation}
    2\pi\bar{x}_{\mathrm{p}} + \eta_{\mathrm{p}}\sin(2\pi
    \bar{x}_{\mathrm{p}})=\pi/2.
    \label{eq:def_xp}
\end{equation}
In this case, $\max{W_{\mathrm{p}}' (\bar{x})} = 2\pi(1-
\eta_{\mathrm{p}})$, $\min{W_{\mathrm{p}}' (\bar{x})} =
-2\pi(1+\eta_{\mathrm{p}})$, $f_{\mathrm{0p}} =
2\pi(1-\eta_{\mathrm{p}})V_{\mathrm{0p}}/d$,
{\setlength\arraycolsep{2pt}
\begin{eqnarray}
    G_{\mathrm{p}}(\bar{x}) &=& \cos{\{2\pi(\bar{x}+\bar{x}_
    {\mathrm{p}}) + \eta_{\mathrm{p}}\sin[2\pi(\bar{x}+\bar{x}_
    {\mathrm{p}})]\}} \nonumber\\[6pt]
    && \times \{1+ \eta_{\mathrm{p}}\cos[2\pi(\bar{x}+\bar{x}_
    {\mathrm{p}})]\}(1-\eta_{\mathrm{p}})^{-1}, \qquad
    \label{eq:def_Gp}
\end{eqnarray}
$G_{\mathrm{p}} (\bar{x}_{1})=-1$ if $\bar{x}_{1} = 1/2 -
\bar{x}_{\mathrm{p}}$, $G_{\mathrm{p}} (\bar{x}_{2}) =
G_{\mathrm{p}}(0) = 0$ if $\bar{x}_{2} = 1 - 2\bar{x}_
{\mathrm{p}}$, and $G_{\mathrm{p}} (\bar{x}_{4}) = \max
{G_{\mathrm{p}}(\bar{x})} = (1 + \eta_{\mathrm{p}})/(1 -
\eta_{\mathrm{p}})$ if $\bar{x}_{4} = 1-\bar{x}_{\mathrm{p}}$.

\cleardoublepage

\begin{figure}
    \centering
    \includegraphics{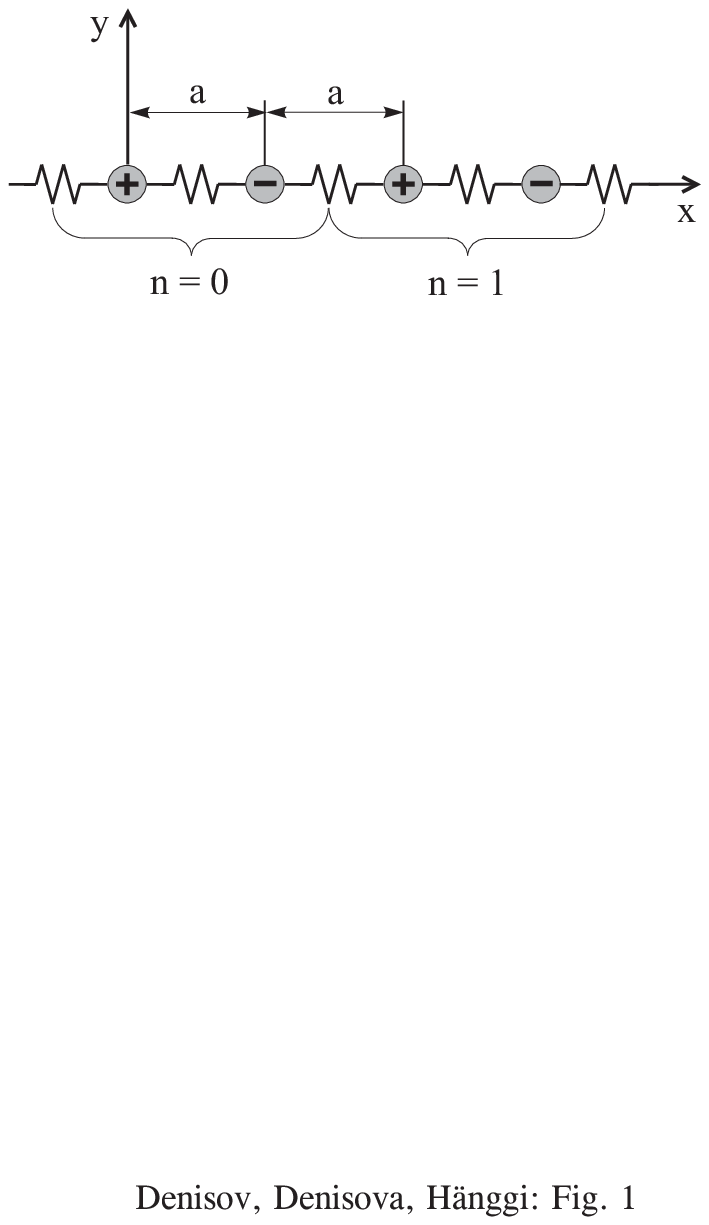}
    \caption{Schematic representation of the ionic chain.}
    \label{fig1}
\end{figure}

\cleardoublepage

\begin{figure}
    \centering
    \includegraphics{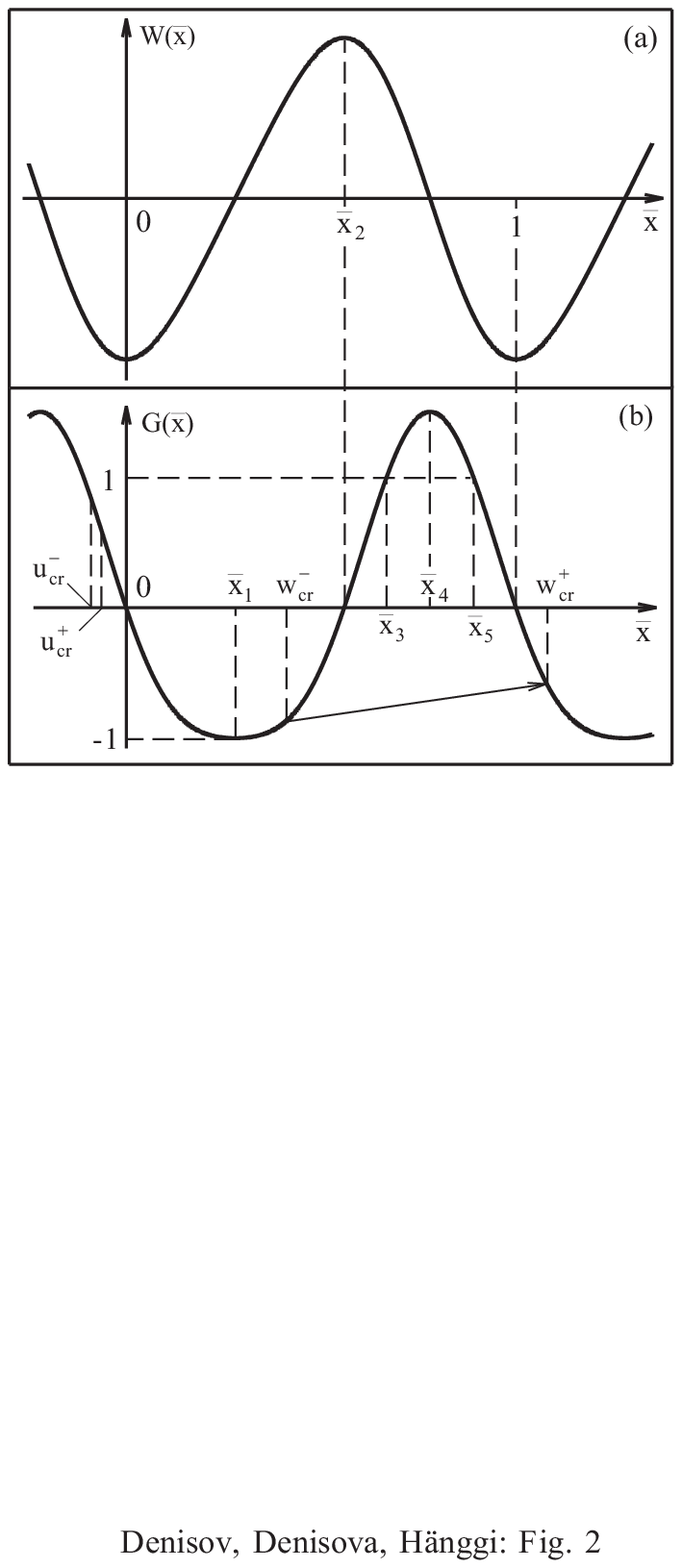}
    \caption{Form of the reduced ratchet potential
    $W(\bar{x})$, part (a), and the corresponding force field
    $G(\bar{x})$, part (b), that are under consideration in this
    work.}
    \label{fig2}
\end{figure}

\cleardoublepage

\begin{figure}
    \centering
    \includegraphics{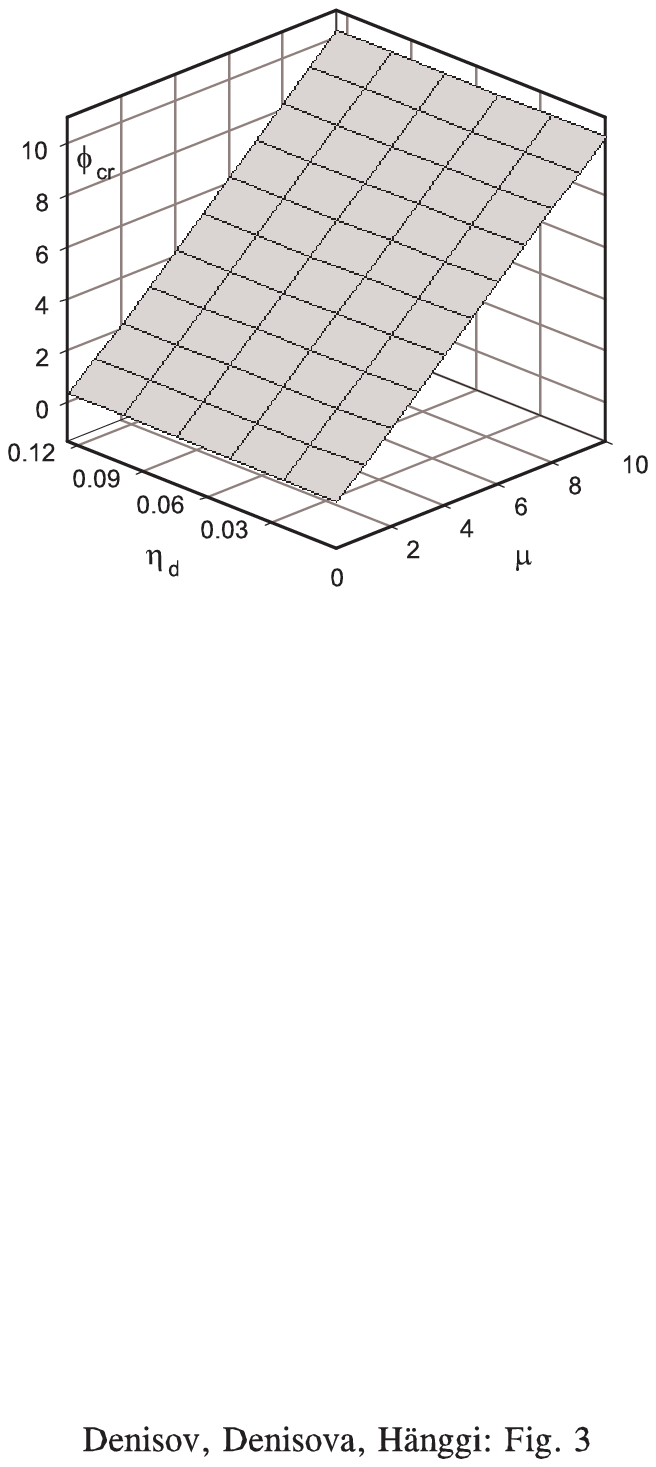}
    \caption{3-D plot of $\phi_{\mathrm{cr}}$ as a function of
    $\mu$ and $\eta_{\mathrm{d}}$.}
    \label{fig3}
\end{figure}

\cleardoublepage

\begin{figure}
    \centering
    \includegraphics{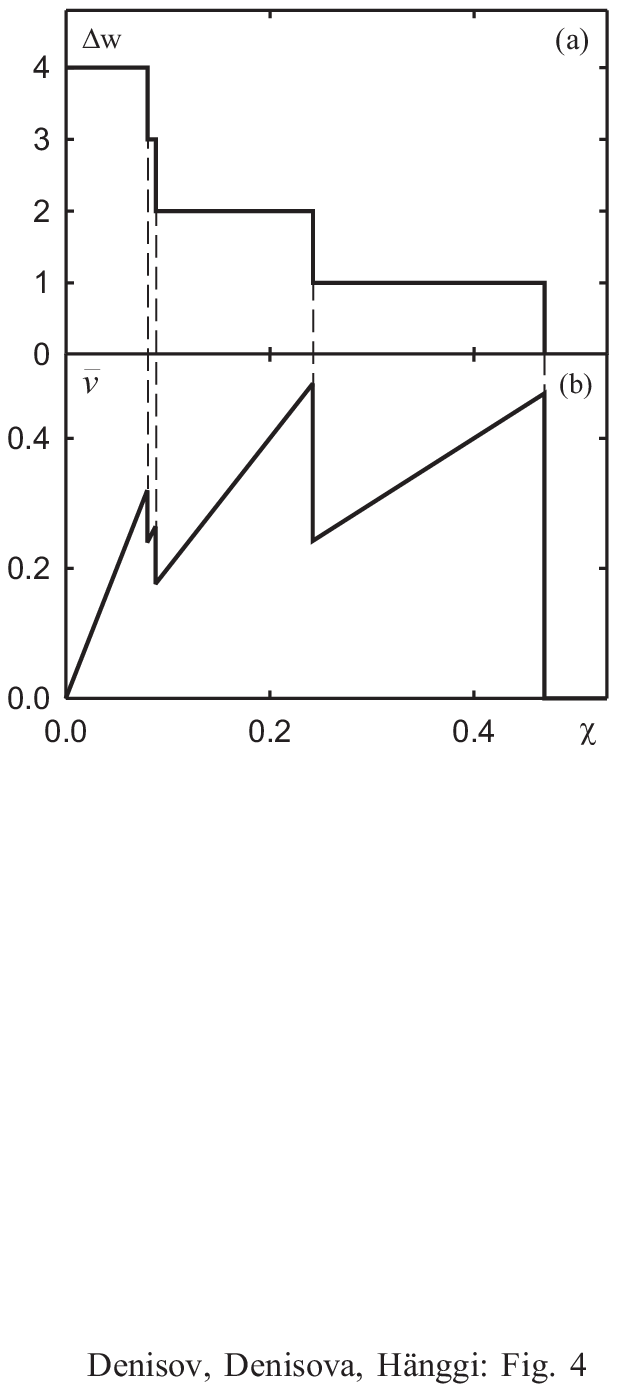}
    \caption{Plots of $\Delta w$, part (a), and $\bar{v}$,
    part (b), vs $\chi$ for $\phi=3$ and $\mu=2$. The intervals
    $(\chi_{1}^{s}, \chi_{2}^{s})$ ($s=1,...,4$) are not
    visible on this scale.}
    \label{fig4}
\end{figure}

\cleardoublepage

\begin{figure}
    \centering
    \includegraphics{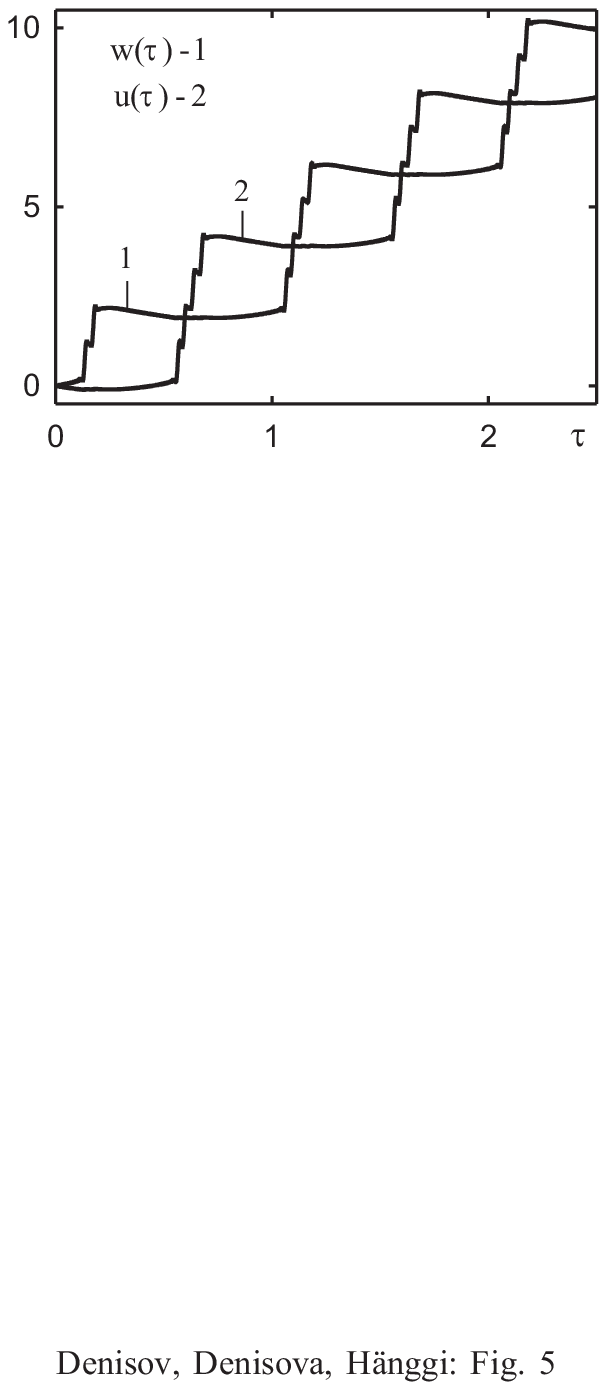}
    \caption{Plots of the particle displacements $w(\tau)$ and
    $u(\tau)$ for $\phi=3$, $\mu=2$, and $\chi=10^{-4}$.}
    \label{fig5}
\end{figure}

\cleardoublepage

\begin{figure}
    \centering
    \includegraphics{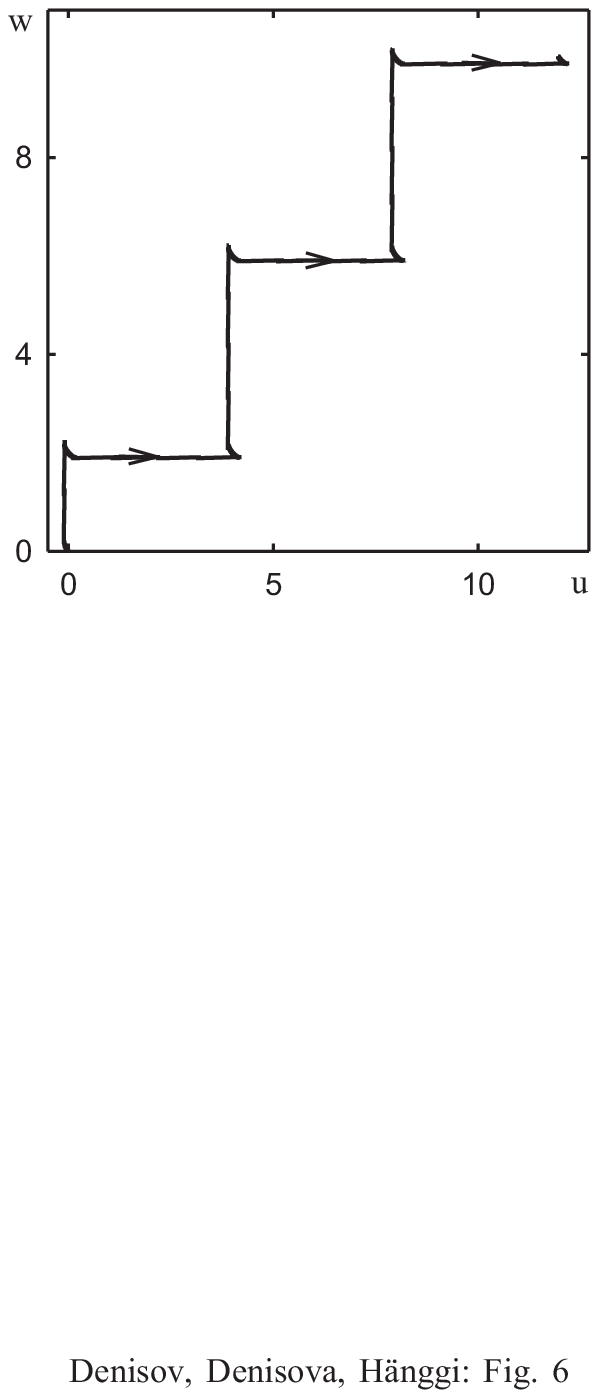}
    \caption{Phase trajectory of the chain motion for
    the same parameters as in Fig.~\ref{fig5}.}
    \label{fig6}
\end{figure}

\cleardoublepage

\begin{figure}
    \centering
    \includegraphics{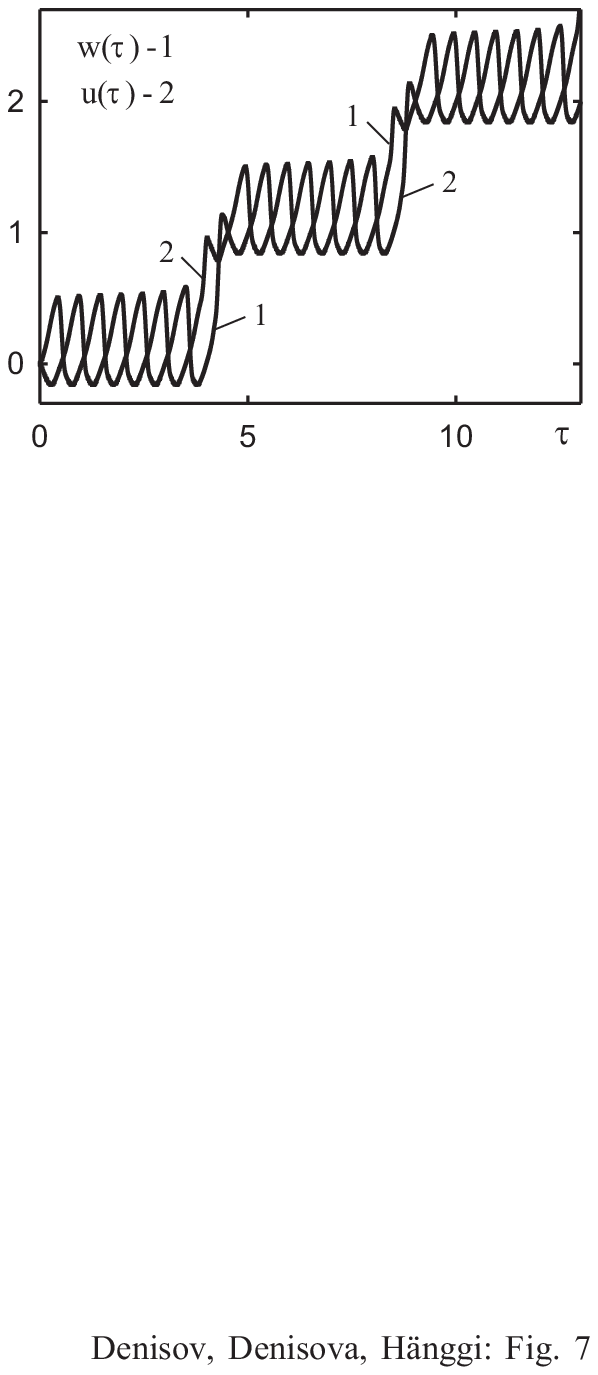}
    \caption{Plots of the particle displacements $w(\tau)$ and
    $u(\tau)$ for $\phi=3$, $\mu=2$, and $\chi=0.473$.}
    \label{fig7}
\end{figure}

\cleardoublepage

\begin{figure}
    \centering
    \includegraphics{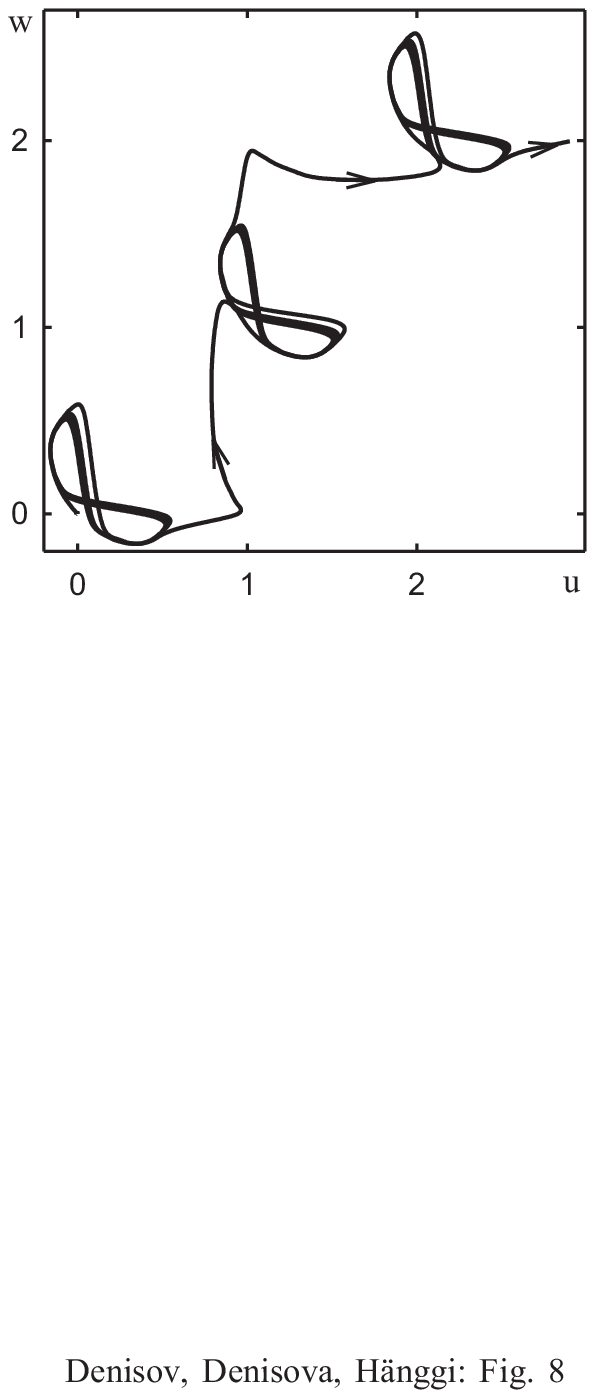}
    \caption{Phase trajectory of the chain motion for
    the same parameters as in Fig.~\ref{fig7}.}
    \label{fig8}
\end{figure}

\begin{thebibliography} {99}
\bibitem{R02} P.~Reimann, Phys.\ Rep.\ \textbf{361}, 57 (2002);
P.~Reimann and P.~H\"{a}nggi, Appl.\ Phys.\ A \textbf{75}, 169
(2002).
\bibitem{Linke02} Special issue on \textit{Ratchets and Brownian
motors: Basics, experiments and applications}, edited by H.~Linke
[Appl.\ Phys.\ A \textbf{75}, 167 (2002)].
\bibitem{JAP97} F.~J\"{u}licher, A.~Ajdari, and J.~Prost, Rev.\
Mod.\ Phys.\ \textbf{69}, 1269 (1997).
\bibitem{A97} R.D.~Astumian, Science\ \textbf{276}, 917 (1997).
\bibitem{AH02} R.D.~Astumian and P.~H\"{a}nggi, Physics\ Today\
\textbf{55} (11), 33 (2002).
\bibitem{RSAP94} J.~Rousselet, L.~Salome, A.~Ajdari, and
J.~Prost, Nature\ (London)\ \textbf{370}, 446 (1994).
\bibitem{FBKL95} L.P.~Faucheux, L.S.~Bourdieu, P.D.~Kaplan,
and A.J.~Li\-b\-chaber, Phys.\ Rev.\ Lett.\ \textbf{74}, 1504
(1995).
\bibitem{LJDB99} C.-S.~Lee, B.~Jank\'{o}, I.~Der\'{e}nyi, and
A.-L.~Barab\'{a}si, Nature\ (London)\ \textbf{400}, 337 (1999).
\bibitem{Kehr1997} K.W.~Kehr, K.~Mussawisade, T.~Wichmann,
and W.~Dieterich, Phys.\ Rev.\ E\ \textbf{56}, R2351 (1997).
\bibitem{DLB98} I.~Der\'{e}nyi, C.~Lee, and A.-L.~Barab\'{a}si,
Phys.\ Rev.\ Lett.\ \textbf{80}, 1473 (1998).
\bibitem{Miretartes} R.~Guantes and S.~Miret-Art\'{e}s, Phys.\ Rev.\
E\ \textbf{67}, 046212 (2003); S.~Sengupta, R.~Guantes,
S.~Miret-Art\'{e}s, and P.~H\"{a}nggi, Physica\ A\ \textbf{338},
406 (2004).
\bibitem{Bartussek1994} R.~Bartussek, P.~H\"{a}nggi, and J.G.~Kissner,
Europhys.\ Lett.\ \textbf{28}, 459 (1994).
\bibitem{HTB} P.~H\"{a}nggi, P.~Talkner, and M.~Borkovec, Rev.\ Mod.\
Phys.\ \textbf{62}, 251 (1990).
\bibitem{M93} M.O.~Magnasco, Phys.\ Rev.\ Lett.\ \textbf{71},
1477 (1993).
\bibitem{Bartussek1996} R.~Bartussek, P.~Reimann, and P.~H\"{a}nggi,
Phys.\ Rev.\ Lett.\ \textbf{76}, 1166 (1996).
\bibitem{CFV97} Z.~Csah\'{o}k, F.~Family, and T.~Vicsek, Phys.\
Rev.\ E\ \textbf{55}, 5179 (1997).
\bibitem{ZHH98} Z.~Zheng, B.~Hu, and G.~Hu, Phys.\ Rev.\ E\
\textbf{58}, 7085 (1998).
\bibitem{Wambaugh1999} J.F.~Wambaugh, C.~Reichhardt, C.J.~Olson,
F.~Marchesoni, and F.~Nori, Phys.\ Rev.\ Lett.\ \textbf{83}, 5106
(1999).
\bibitem{Savelev2004} S.~Savel'ev, F.~Marchesoni, and F.~Nori,
Phys.\ Rev.\ Lett.\ \textbf{92}, 160602 (2004).
\bibitem{JKH96} P.~Jung, J.G.~Kissner, and P.~H\"{a}nggi, Phys.\
Rev.\ Lett.\ \textbf{76}, 3436 (1996).
\bibitem{SL99} A.~Sarmiento and H.~Larralde, Phys.\ Rev.\ E\
\textbf{59}, 4878 (1999).
\bibitem{M00} J.L.~Mateos, Phys.\ Rev.\ Lett.\ \textbf{84},
258 (2000).
\bibitem{S01} I.M.~Sokolov, Phys.\ Rev.\ E\ \textbf{63},
021107 (2001).
\bibitem{BCM02} M.~Borromeo, G.~Costantini, and F.~Marchesoni,
Phys.\ Rev.\ E\ \textbf{65}, 041110 (2002).
\bibitem{Marchesoni1996} F.~Marchesoni, Phys.\ Rev.\ Lett.\
\textbf{77}, 2364 (1996).
\bibitem{Hirahara00} K.~Hirahara, K.~Suenaga, S.~Bandow, H.~Kato,
T.~Oka\-zaki, H.~Shinohara, and S.~Iijima, Phys.\ Rev.\ Lett.\
\textbf{85}, 5384 (2000).
\bibitem{SF02} Z.~Siwy and A.~Fuli\'{n}ski, Phys.\ Rev.\ Lett.\
\textbf{89}, 198103 (2002).
\bibitem{Reimann1999} P.~Reimann, R.~Kawai, C.~Van~den~Broeck,
and P.~H\"{a}n\-ggi, Europhys. Lett. \textbf{45}, 545 (1999).
\bibitem{DD03} S.I.~Denisov and E.S.~Denisova, Phys.\ Rev.\ B\
\textbf{68}, 064301 (2003).
\bibitem{Moon} F.C.~Moon, \textit{Chaotic Vibrations}
(Wiley, New York, 1987).
\end{thebibliography}
\end{document}